\documentstyle[aps,twocolumn]{revtex}
\draft

\begin{document}
\title{Quantum Teleportation Using Quantum Non-Demolition Technique}
\author{D. B. Horoshko\cite{email} and S. Ya. Kilin}
\address{Institute of Physics, Belarus National Academy of Sciences, 
Minsk 220072 Belarus}
\maketitle

\begin{abstract}
We propose a new scheme and protocol for quantum teleportation of a
single-mode field state, based on entanglement produced by quantum
non-demolition interaction. We show that the recently attained results in
QND technique allow to perform the teleportation in quantum regime. We also
show that applying QND coupling to squeezed fields will significantly
improve the quality of teleportation for a given degree of squeezing.
\end{abstract}


Quantum teleportation is a transport of quantum state of a system to another
similar system via a classical channel \cite{Ben}. The principles of quantum
mechanics demand that for realization of such a transport, the sender and
the receiver must have two ancillary quantum systems, being strongly
quantum-mechanically correlated (entangled). Quantum teleportation is a
fundamental phenomenon of quantum world and also one of the key procedures
in the rapidly growing area of quantum information \cite{QI}. Recently
several teleportation schemes have been proposed \cite{tele-theory} and the
successful realization of some of them has been reported: teleportation of
the polarization state of a single photon \cite{Bou} \cite{Bosc}, of the
state of a single-mode optical field \cite{Fur}, \cite{Lam} and of nuclear
spin state \cite{Lafl}. In this Letter we propose a new scheme for quantum
teleportation of the state of single-mode optical field. In contrast to the
existing schemes \cite{Fur}, \cite{Lam}, where a superposition of two
squeezed fields has been used as a source of entanglement, we propose to use
the entanglement produced by quantum non-demolition (QND) interaction \cite
{Brag}, \cite{Yurke}, realized in the last years for optical fields in
various media \cite{Grangier}, the best results being obtained for coupling
two optical beams by parametric interaction in a $\chi ^{(2)}$ crystal \cite
{crystal-qnd} or using cold atoms in a trap \cite{trap-qnd}.

Our scheme is depicted in Fig. 1. The nonlinear QND interaction produces the
following coupling between two bright coherent fields $E_a$ and $E_b$ \cite
{Yurke} 
\begin{eqnarray}
\tilde{E}_a &=&E_a+\frac g2\left( E_b^{+}+E_b\right) ,  \label{1} \\
\tilde{E}_b &=&E_b+\frac g2\left( E_a^{+}-E_a\right) ,  \label{2}
\end{eqnarray}
where $\tilde{E}_a$ and $\tilde{E}_b$ are the outgoing fields and $g$ is the
QND gain. It is easy to find that the quadratures $X=E^{+}+E$ and $%
Y=i(E^{+}-E)$ of both fields are transformed as 
\[
\begin{array}{ll}
\tilde{X}_a=X_a+gX_b, & \tilde{Y}_a=Y_a, \\ 
\tilde{X}_b=X_b, & \tilde{Y}_b=Y_b-gY_a.
\end{array}
\]
The quadratures $X_b$ and $Y_a$ are the so-called QND variables, which are
not affected by coupling. The conjugated quadratures $X_a$ and $Y_b$ become
correlated with $X_b$ and $Y_a$ respectively and for infinitely large QND
gain, or for infinitely squeezed $Y_b$ ($X_a$), the measurement of $Y_b$ ($%
X_a$) gives a result proportional to that of measurement of $Y_a$ ($X_b$).
This is the main idea of QND measurement. In our approach this measurement
is not performed, but the entanglement produced by QND coupling is used for
quantum teleportation, as described below.

One of the resulting from QND interaction beams ($\tilde{E}_a$) passes to
the sender Alice, and another one ($\tilde{E}_b$) passes to the receiver Bob
for further reconstruction of the teleported state. The protocol of quantum
teleportation is as follows. Alice superimposes the field $E_{in}$, which is
to be teleported, with the field $\tilde{E}_a$ at a beam splitter with the
transmittance $\epsilon $, and, using homodyne photodetection, measures the $%
X$ quadrature of the reflected (for $E_{in}$) field and the $Y$ quadrature
of the transmitted field, obtaining the values 
\begin{eqnarray}
x &=&\epsilon ^{1/2}\left( X_a+gX_b\right) +(1-\epsilon )^{1/2}X_{in},
\label{3} \\
y &=&\epsilon ^{1/2}Y_{in}-(1-\epsilon )^{1/2}Y_a,  \label{4}
\end{eqnarray}
where the field variables are written in the Wigner representation. The
measured values are sent to Bob via a classical channel, and upon receiving
them Bob prepares the bright laser beam $E_\beta $ by means of phase and
amplitude modulation in a coherent state with the amplitude $\beta =\Gamma
G(x+igy)/2$, where $\Gamma \gg 1$ and $G$ are some real gains. After that
Bob superimposes the field $E_\beta $ with the beam $\tilde{E}_b$ at a
beam-splitter with very low transmittance $\epsilon ^{\prime }=\Gamma
^{-2}\ll 1$, so that the reflected field $\tilde{E}_b$ is just shifted by $%
G(x+igy)/2$, but no noise is added. The result of interference of the
reflected field $\tilde{E}_b$ and the transmitted field $E_\beta $ is the
output $E_{out}$: 
\begin{eqnarray}
2E_{out} &=&2\Gamma ^{-1}E_\beta -2\tilde{E}_b=G(x+igy)-X_b-iY_b+igY_a 
\nonumber  \label{5} \\
&=&G\epsilon ^{1/2}X_a+(Gg\epsilon ^{1/2}-1)X_b+G(1-\epsilon )^{1/2}X_{in}
\label{5} \\
&&+ig\left( 1-G(1-\epsilon )^{1/2}\right) Y_a-iY_b+iGg\epsilon ^{1/2}Y_{in}.
\nonumber
\end{eqnarray}
Now, if the electric gain $G$ and the transmittance $\epsilon $ satisfy for
a given QND gain $g$ the relations $G=(1-\epsilon )^{-1/2}$, $g=(1-\epsilon
)^{1/2}\epsilon ^{-1/2}$, the output field reads as 
\begin{equation}
2E_{out}=X_{in}+iY_{in}+g^{-1}X_a-iY_b,  \label{7}
\end{equation}
that is, for $\langle X_a\rangle =\langle Y_b\rangle =0$, it reproduces the
incoming field with additional noise caused by quadratures $X_a$ and $Y_b$.
For infinite QND gain $g$ and infinitely squeezed quadrature $Y_b$ the
teleportation is ideal. However, even for finite QND gain and not squeezed $%
Y_b$ the proposed teleportation scheme works in an essentially quantum
regime.

This can be seen from comparing this scheme to classical teleportation. In
the latter case Alice and Bob do not have correlated ancillary systems and
all they can do in order to accomplish a true reproduction of the average
values of $X_{in}$ and $Y_{in}$, is to measure this quadratures with
uncertainty imposed by the Heisenberg's principle and create a coherent
field with the average values of quadratures equal to measured ones. So the
protocol of classical teleportation is as follows. Alice divides the
incoming field $E_{in}$ in two parts by a 50:50 beam splitter and measures
the $X$ quadrature of the reflected field and the $Y$ quadrature of the
transmitted field, obtaining the values 
\begin{eqnarray}
x &=&\frac 1{\sqrt{2}}X_{in}+\frac 1{\sqrt{2}}X_v,  \label{8} \\
y &=&\frac 1{\sqrt{2}}Y_{in}-\frac 1{\sqrt{2}}Y_v,  \label{9}
\end{eqnarray}
where $E_v=\left( X_v+iY_v\right) /2$ is the vacuum field at the unused port
of the beam-splitter. The measured values are send to Bob who creates a
coherent state 
\begin{eqnarray}
2E_{out} &=&\sqrt{2}(x+iy)+2E_w  \label{10} \\
&=&X_{in}+iY_{in}+X_v-iY_v+X_w+iY_w,  \nonumber
\end{eqnarray}
where $E_w=\left( X_w+iY_w\right) /2$ is the vacuum fluctuation inevitably
added at the reconstruction stage (e.g., as in the previous protocol, Bob
can superimpose a strong coherent field $2E_\beta =\Gamma \sqrt{2}(x+iy)$
with vacuum field $E_w$ at a low transmittance $\epsilon ^{\prime }=\Gamma
^{-2}$ beam splitter and use the reflected (for $E_w$) field as output).
Now, to compare quantum (Eq. (\ref{7})) and classical (Eq. (\ref{10}))
teleportation, we must calculate a quantitative measure of teleportation
success, connected with some verification procedure. Let us consider the
verification scheme shown in Fig. 2, similar to that of Ref. \cite{Fur}. The
verifier Victor, independent on Alice and Bob, prepares an ensemble of
states of field $E_0$. Victor can directly measure the average value and the
variance of the field quadrature $X_0$. Or he can shift the phase of the
field by random phase $\varphi (t)$ and give it to Alice who teleports the
state to Bob. After receiving the state $E_{out}$ from Bob, Victor again
shifts the phase, this time by $-\varphi (t)$ and measures the $X$%
-quadrature of the obtained field $E_T$. If the teleportation is ideal, the
average value and variance of $X_T$ will be the same as for $X_0$. The
proposed in this Letter teleportation scheme gives the following expression
for $X_T$: 
\begin{equation}
X_T=X_0+g^{-1}X_a\cos \varphi -Y_b\sin \varphi .  \label{11}
\end{equation}
So for $\langle X_a\rangle =\langle Y_b\rangle =0$ the average value of $X_T$
is equal to that of $X_0$. The variance of $X_T$ is given by 
\begin{eqnarray}
\left\langle \left( \delta X_T\right) ^2\right\rangle &=&\left\langle \left(
\delta X_0\right) ^2\right\rangle +g^{-2}\left\langle \left( \delta
X_a\right) ^2\right\rangle \overline{\cos ^2\varphi }  \label{12} \\
&&+\left\langle \left( \delta Y_b\right) ^2\right\rangle \overline{\sin
^2\varphi },  \nonumber
\end{eqnarray}
where $\delta X=X-\langle X\rangle $ and the bar denotes averaging over the
random phase shift. If the phase is distributed homogeneously in $[0,2\pi ]$%
, then $\overline{\cos ^2\varphi }=\overline{\sin ^2\varphi }=\frac 12$, and 
\begin{eqnarray}
N_{add} &\equiv &\left\langle \left( \delta X_T\right) ^2\right\rangle
-\left\langle \left( \delta X_0\right) ^2\right\rangle  \label{13} \\
&=&\frac 1{2g^2}\left\langle \left( \delta X_a\right) ^2\right\rangle +\frac 
12\left\langle \left( \delta Y_b\right) ^2\right\rangle ,  \nonumber
\end{eqnarray}
where $N_{add}$ is the noise added to the teleported light. We use the added
noise as a measure of teleportation success instead of commonly used
fidelity $F=\langle \psi _{in}|\rho _{out}|\psi _{in}\rangle $ (where $|\psi
_{in}\rangle $ is the incoming state and $\rho _{out}$ is the density matrix
of the output field), for two following reasons. Firstly, fidelity generally
depends on the incoming state, while for the teleportation of field state,
which is a linear in field transformation, the output field is always the
incoming one plus some noise; so it is natural to describe the teleportation
by this added noise, independent on the incoming state. Secondly, the output
state can differ from the input one in two ways: in averages (displacement)
and in variances of quadratures (distortion). Fidelity (as well as the
parameters introduced in Ref. \cite{Lam}) takes into account both these
effects. However, classical teleportation always can be arranged so that the
average values (of quadratures) are transported correctly. So only a measure
of state distortion is necessary for characterizing the quantum nature of
teleportation, and the added noise just provides such a measure.

It is easy to find from Eq. (\ref{10}) that for classical teleportation 
\begin{eqnarray*}
N_{add}^{cl} &=&\frac 12\left\langle \left( \delta X_v\right)
^2\right\rangle +\frac 12\left\langle \left( \delta Y_v\right)
^2\right\rangle +\frac 12\left\langle \left( \delta X_w\right)
^2\right\rangle \\
+\frac 12\left\langle \left( \delta Y_w\right) ^2\right\rangle &\geq &2
\end{eqnarray*}
for any states of fields $E_v$ and $E_w$, which can be seen from the
inequality $a^2+b^2\geq 2ab$ and the Heisenberg's uncertainty principle. So,
to reach the quantum regime of teleportation, one must reach added noise $%
0\leq N_{add}<2$, under condition that $\langle X_T\rangle =\langle
X_0\rangle $ for any phase $\varphi $. In the proposed scheme it can be
obtained by using coherent fields $E_a$ and $E_b$ (with imaginary and real
amplitudes respectively) and QND gain $g>1/\sqrt{3}$.

The entanglement produced by a QND device (or its quantum state preparation
regime) is usually characterized by conditional variance of signal given a
measured value of meter, $V_c$ \cite{Walls}. For transformation described by
Eqs. (\ref{1}), (\ref{2}) and coherent input fields $E_a$ and $E_b$ this
variance is given by 
\begin{equation}
V_c\equiv \left\langle \left( \delta \tilde{X}_b\right) ^2\right\rangle -%
\frac{|\left\langle \delta \tilde{X}_a\delta \tilde{X}_b\right\rangle |^2}{%
\left\langle \left( \delta \tilde{X}_a\right) ^2\right\rangle }=\frac 1{1+g^2%
}  \label{15}
\end{equation}
and the condition $g>1/\sqrt{3}$ corresponds to the condition $V_c<\frac 34$%
. In a recent QND experiment using cold atoms in a trap as a nonlinear
medium \cite{trap-qnd} the value $V_c=0.45$ has been measured ($V_c=0.37$
with corrections for optical losses). In the same time, our requirements for
field phases are exactly that needed for QND interaction based on
cross-phase modulation, implemented in this experiment. For the QND schemes
based on $\chi ^{(2)}$ crystals, a value $V_c=0.65-0.7$ has been achieved 
\cite{crystal-qnd}, without restrictions on the phases of interacting beams.
These results speak about a very good possibility to reach the quantum
regime of teleportation using the existing experimental technique and the
protocol presented in this Letter.

Let us briefly compare our scheme to that using squeezed states \cite{Fur}, 
\cite{Lam}. In classical teleportation the output field is a sum of the
input one and four additional noises (quantum duties), as shown by Eq. (\ref
{10}), and the noise added to a random quadrature is always more than 2.
Quantum teleportation can be realized by using correlated ancillary fields $%
E_v$ and $E_w$, produced by mixing two squeezed fields $E_1$ and $E_2$ at a
50:50 beam-splitter, so that in Eq. (\ref{10}) $X_v+X_w=\sqrt{2}X_1$ and $%
Y_v-Y_w=\sqrt{2}Y_2$, and the added noise is given by 
\begin{equation}
N_{add}^{sq}=V_1+V_2,  \label{16}
\end{equation}
where $V_1$ and $V_2$ are variances of quadratures $X_1$ and $Y_2$
respectively. When the quadratures $X_1$ and $Y_2$ are squeezed, the added
noise becomes less than the classical value of 2. In our approach the
ancillary fields are correlated via a QND interaction. Two QND variables ($%
X_b$ and $Y_a$) are automatically cancelled in the output field due to QND
correlations. One of the remaining noises ($X_a$) can be made as small as
desired by increasing the QND gain and/or by squeezing this quadrature. The
noise of $Y_b$ can be reduced by squeezing this quadrature only. The added
noise in this case reads as ($V_c<1$) 
\begin{equation}
N_{add}^{QND}=\frac 12\frac{V_c}{1-V_c}V_a+\frac 12V_b,  \label{17}
\end{equation}
where $V_a$ and $V_b$ are variances of quadratures $X_a$ and $Y_b$
respectively. Comparing Eq. (\ref{17}) with Eq. (\ref{16}) we see that with
decreasing conditional variance the QND coupling decreases significantly the
requirements for the degree of squeezing of ancillary beams (so that for $%
V_c<\frac 34$ they can be not squeezed at all).

It should be noted that a principal possibility to use QND coupling for
quantum teleportation has been pointed in Ref. \cite{Milburn}. However this
paper considers only QND gain $g=1$ and two ideally squeezed ancillary
beams, in which case no benefits from QND coupling can be seen, compared to
mixing these beams at a beam-splitter.

In summary, we have described a protocol for quantum teleportation of a
single mode optical field using QND coupling as a source of entanglement of
ancillary laser beams. In our scheme the QND device is used in a quantum
state preparation regime, and the recent achievements for realizing this
regime satisfy very well the requirements for quantum teleportation. From
the other side, the existing schemes for quantum teleportation of
single-mode field state are limited by the restricted degree of squeezing,
available at present, and using QND coupling for squeezed ancillary beams
can provide further improvement of quantum teleportation technique.

This work was supported by INTAS, grant \# 96-0167.

\begin{figure}[tbp]
\caption{Schematic of quantum teleportation arrangement. Two entangled beams 
$\tilde{E}_a$ and $\tilde{E}_b$ are produced by the QND device. The incoming
field $E_{in}$ is mixed by Alice with one of the entangled beams and the
quadratures of the resulting fields are measured by two homodyne detectors,
using purely real local oscillator field ${\cal E}_{LO}$ and a purely
imaginary one $i{\cal E}_{LO}$. The measurement results are sent via a
classical channel to Bob who prepares a laser beam in a proper coherent
state by means of amplitude and phase modulator $M$. This coherent field is
superimposed with the second entangled beam, resulting in the output field.
The state of the latter reproduces closely the state of the incoming field.}
\label{figure1}
\end{figure}

\begin{figure}[tbp]
\caption{The verification procedure. The verifier, completely independent on
Alice and Bob, prepares an ensemble of states for field $E_0$. This field
can be directly sent to the homodyne detector (dashed line) for measuring
the distribution of its real quadrature. Either it can be sent to Alice for
teleportation with the following measurement of the field received from Bob
(solid line). The difference in noise level between two paths is a measure
of the teleportation success. The phase shifts, introduced by phase
modulators $PM$ allow to control the noise added to a corresponding
quadrature of the teleported field.}
\label{figure2}
\end{figure}

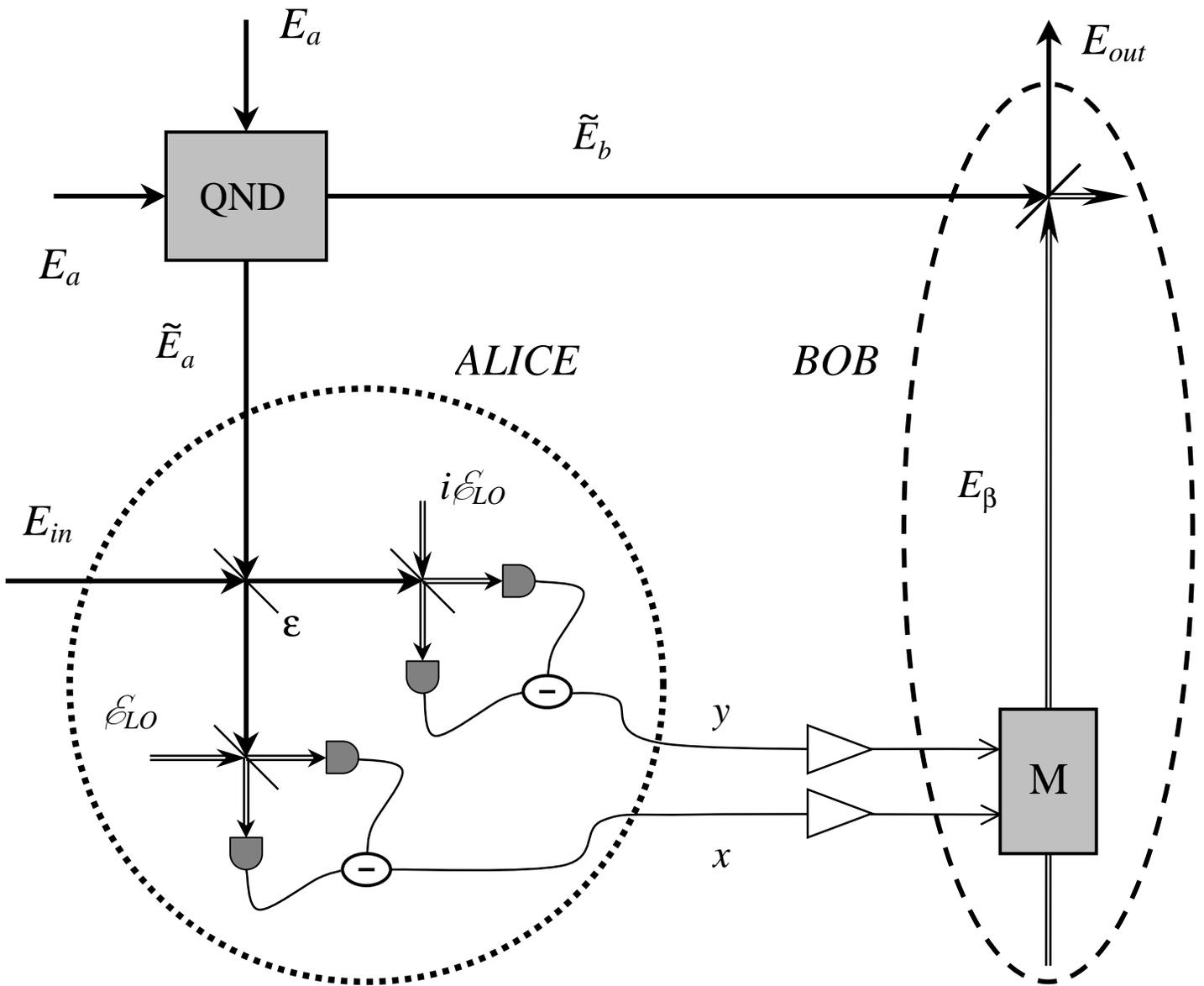
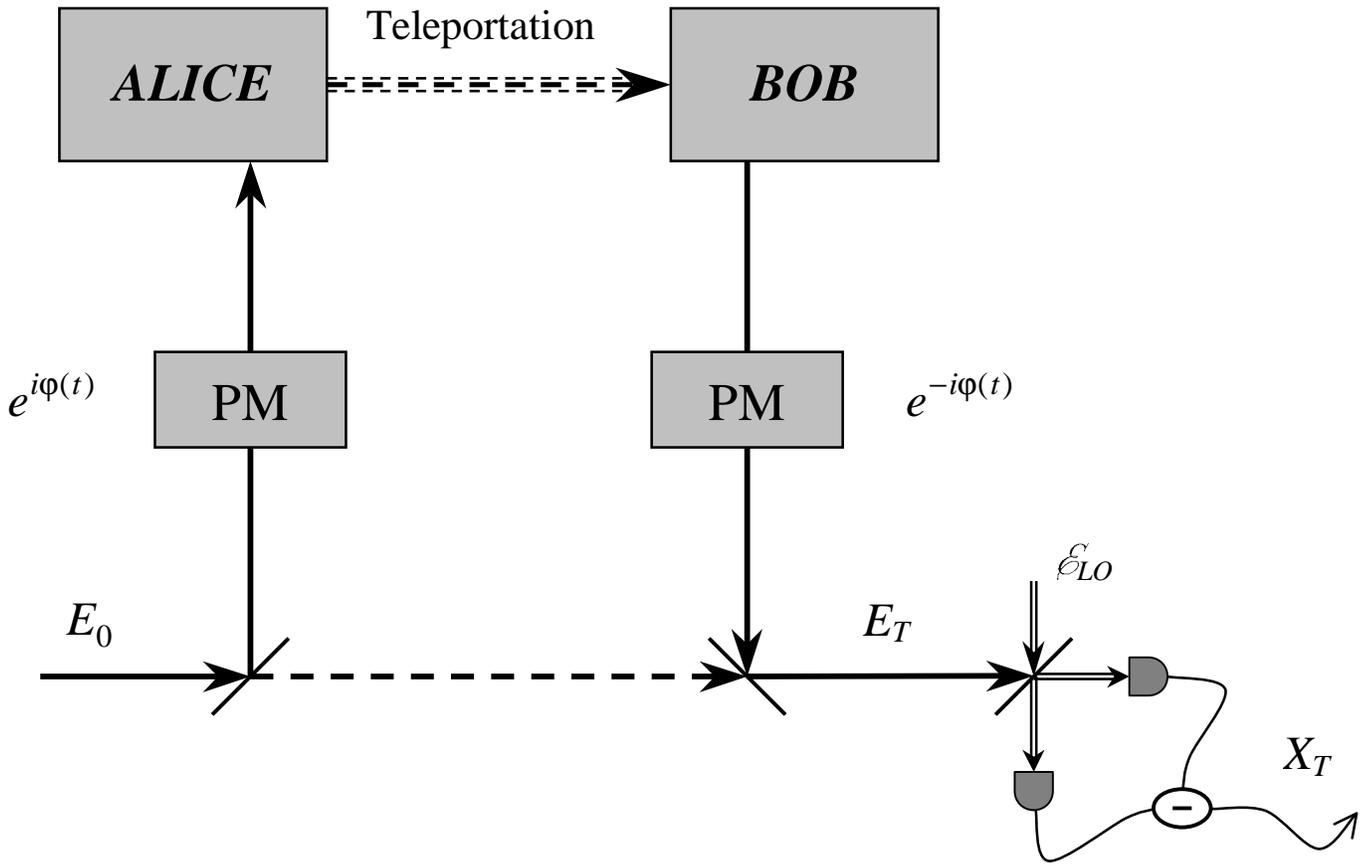
\end{document}